\documentclass[12pt]{iopart}
\usepackage[english]{babel}
\usepackage[latin1]{inputenc}
\usepackage{iopams,setstack}
\usepackage{graphicx}

\begin{document}

\title[Quantum transport of non-interacting Fermi gas]{Quantum transport of non-interacting Fermi gas in an optical lattice combined with harmonic trapping}
\author{V Ruuska and P T\"orm\"a}
\address{Department of Physics, NanoScience Center, P.O. Box 35 (YFL),
       FIN-40014 University of Jyv\"askyl\"a}

\begin{abstract}
 We consider a non-interacting Fermi gas in a combined harmonic and periodic potential. We calculate
the energy spectrum and simulate the motion of the gas after sudden replacement of the trap center. 
For different parameter regimes, the system presents dipole oscillations, damped oscillations around the replaced
center, and localization. The behaviour is explained by the change of the energy spectrum from linear to quadratic.
\end{abstract}


\section{Introduction}
 
In solid state physics, electrons in an atomic lattice are modelled by
fermions satisfying the Schrödinger equation in a periodic potential.
Alkali atoms in a periodic potential created by light have
recently provided a new physical realization of the same mathematical
model \cite{bose1,bose2,modugno1,modugno2,Pezze,Roati}. Experimentally, optical lattices are highly controllable and
open a wide range of parameters for study, also the dynamics. For bosonic atoms, quantum transport such as Bloch 
oscillations have been observed \cite{bose2}. The first such experiments on fermions in optical lattices 
have been realized very recently \cite{modugno1,modugno2,Pezze,Roati}. 

Unlike electrons in bulk solids, the atomic gases feel in addition to 
the lattice an overall confining potential, usually of harmonic shape. This harmonic trapping can be chosen 
weak and the system considered nearly homogeneous. On the other hand, it can be made significant in order to
induce interesting quantum transport experiments: shift of the trap center may cause oscillatory motion of
the gas \cite{modugno1,modugno2,Pezze}. Single-particle states and spectrum in a combined harmonic and 
periodic potential have been studied in \cite{Rigol, Hooley}, and the effect of the potential on quantum 
transport has been analyzed in \cite{Pezze,yksi} using semiclassical analysis.
We present here exact numerical treatment which gives a complementary point of view
for understanding the problem, and
provides a method applicable for a wide range of parameter values.  
Since we consider non-interacting fermions, the only physical assumptions we need are that the
individual atoms obey the Schr\"odinger equation and that 
the equilibrium of the many-particle system can be described by
the grand canonical ensemble. The physical model is considered in
more detail in Section~\ref{physical}. Details of the 
numerical implementation are discussed in Section~\ref{numerical}. 
Finally, in Section~\ref{simulation} the results of the
simulations are presented and discussed, and compared with the experiments. 

\section{Physical model} \label{physical}

We consider a one-dimensional optical lattice combined with a three-dimensional
quadratic potential. The quadratic potential is weak in the direction of the
lattice (axial direction) and tight in the remaining two (radial) directions.
The full three-dimensional Hamiltonian operator for a single particle is 
\begin{equation*}
H^{3D}=-\frac {\hbar^2}{2m} \nabla^2 + \frac{1}{2}m(\omega_a^2x^2+\omega_r^2y^2+\omega_r^2z^2) 
       + \frac{U}{2}\cos (\frac{4\pi}{\lambda}x)
\end{equation*}
where $\omega_a$ and $\omega_r$ are the axial and radial frequencies, respectively, $U$
is the lattice height, and $\lambda$ is the lattice wavelength. 
We shall consider the one-dimensional system in the axial direction first and deal with 
the radial degrees of freedom later.
Choosing the units of energy and length to be $\hbar\omega_a$ and $\sqrt{\hbar/m\omega_a}$, 
respectively,
and using the notation  $A=U/2$ and $k=4\pi/\lambda$, the 
one-dimensional Hamiltonian can be written as a sum 
\begin{equation*}
H= H^O+H^L=-\frac {1}{2} \frac{d^2}{dx^2} + \frac{1}{2}x^2 + A\cos (kx)
\end{equation*}
of the lattice potential $H^L= A\cos (kx)$ and the oscillator Hamiltonian 
\begin{equation*}
H^O=\frac {1}{2} \frac{d^2}{dx^2} + \frac{1}{2}x^2. 
\end{equation*}

Let $(\phi_n)$ and $(\epsilon_n)$ be the eigenfunctions 
and eigenenergies of $H$, respectively.
Using the annihilation and creation operators $a_n, a_n^\dagger$
associated to the eigenfunctions we may write the many-particle Hamiltonian
in the fermionic Fock space as 
\begin{equation*}
\hat{H} = \sum_{n} \epsilon_n a_n^\dagger a_n.
\end{equation*}
Initially the system is in the statistical equilibrium and can be
described by the grand canonical ensemble
\begin{equation}
\Phi(0) = \frac {e^{-\beta (\hat{H}-\mu N)}}{\Tr e^{-\beta (\hat{H}-\mu N)}}
\label{gce}
\end{equation}
where $\mu$ is the chemical potential, $N= \sum_n a_n^\dagger a_n $ is the
number operator, and  $\beta = \frac{1}{k_BT}$. The time evolution of the
system is given by
\begin{equation*}
\Phi(t) = e^{i \hat{H}^d t/\hbar} \Phi(0) e^{-i \hat{H}^d t/\hbar}
\end{equation*} 
where $H^d$ is the Hamiltonian governing the
dynamics of the system. Of course, $H^d$ must be different from the 
Hamiltonian $H$ used to calculate the initial equilibrium state $\Phi(0)$, otherwise the
system will be stable. In the experiment, the displacement $d$ of the harmonic
trap gives rise to the new Hamiltonian
\begin{equation*}
H^d=-\frac {1}{2} \frac{d^2}{dx^2} + \frac{1}{2}(x-d)^2 + A\cos (kx),
\end{equation*}
which starts the time evolution of the system.
Displacement of the trap center in the absense of the lattice
leads to the dipole oscillations characteristic for a harmonic potential.

We want calculate the time evolution of the expectation value of the
position operator. The task is greatly simplified by the fact that
we can represent any many-particle state $\Phi$
by a single-particle state $\rho(\Phi)$ when calculating expectation
values of single-particle observables. This reduction is discussed in more
detail in the Appendix. The single-particle state $\rho(\Phi(0))$ corresponding 
to the grand canonical ensemble $\Phi(0)$ is a diagonal operator 
in the energy eigenbasis. The diagonal elements are given by the Fermi-Dirac distribution 
\begin{equation*}
f(\varepsilon) = \frac {1}{1+e^{\beta \varepsilon}}.
\end{equation*}
When the Hamiltonian operator is itself a single-particle operator, which is the
case for non-interactig Fermi gas, the reduced state obeys the ordinary Schr\"odinger equation.
Hence the final formula for the motion of the Fermi gas is simply
\begin{equation}
\langle X \rangle (t) = \Tr (e^{iH^dt/\hbar} \rho(\Phi(0)) e^{-iH^dt/\hbar} X ),
\label{timeevolution}
\end{equation}
where $X$ is the position operator. 

As the final step we consider dimensional reduction from three dimensions
to one. Even though we are interested in the motion of the Fermi gas 
only in one dimension, the physical system is naturally three-dimensional.
However, we can simply trace out the radial degrees of freedom since
they are independent of the observable which we are measuring. In practice, this 
is done by replacing the Fermi-Dirac distribution by the effective 
one-dimensional energy distribution
\begin{equation*}
f_{eff}(\varepsilon) = \sum_{k,l=0}^\infty f(\varepsilon+(k+l)\hbar\omega_r).
\end{equation*}
where the multiples of $\hbar\omega_r$ exhaust the energy spectra of the
independent oscillators in the radial directions.

\section{Numerical implementation} \label{numerical}

The time evolution of the center of mass of the non-interacting Fermi gas trapped in
an optical lattice was 
calculated using the formula~(\ref{timeevolution}). The first task was to
diagonalize the Hamiltonian. Actually there were two Hamiltonians $H$ and $H^d$ to be 
considered corresponding to the two positions of the trap. However, we
made the simplifying assumption
that the displacement of the trap was a multiple of the lattice wavelength
(this is justified when the latter is much smaller than the former).
Then by a symmetry argument the two sets of eigenfunctions were obtained
from each other by a simple translation.
According to the formula~(\ref{gce}) the initial equilibrium state $\Phi(0)$ is diagonal
in the eigenstate basis of the initial Hamiltonian $H$, and the diagonal elements are
determined by the corresponding eigenvalues. 
The chemical potential $\mu$ was fitted to match the particle number.
By a change of basis, the state matrix was then transformed to
the eigenstate basis of the displaced Hamiltonian $H^d$. The position operator $X$ was
also expressed in this basis.   
By definition, the Hamiltonian itself $H^d$ is diagonal in this basis, so that 
the time evolution was easy to calculate using formula~(\ref{timeevolution}).

The main difficulty was to find a basis suitable for diagonalizing the 
Hamiltonian numerically. We used the Hermite functions
\begin{equation*}
\psi_n(x) = (2^n n! \sqrt{\pi})^{-\frac{1}{2}}e^{-\frac{x²}{2}} H_n(x)
\end{equation*} 
where $H_n$ is the $n$th Hermite polynomial. With this choice the harmonic
oscillator Hamiltonian $H^O$ becomes diagonal with matrix elements
\begin{equation*}
H^O_{nm} = (n+1/2)\delta_{nm}
\end{equation*}
and even the matrix elements of the lattice part $H^L$ can be calculated analytically.
To be exact, the matrix elements are
recovered as $H^L_{nm}=A \mathbf{Re} (I_{nm})$ where 
\begin{equation*}
I_{nm}=I_{nm}(k):=\int_{-\infty}^{\infty} \psi_n(x)\psi_m(x)e^{ikx} \, dx.
\end{equation*}
We were able to derive explicit closed expressions for $I_{nm}(k)$ as
finite polynomials of $k$. However, they are rather cumbersome and not very
useful in practice. In numerical calculations the integrals are most easily obtained
from the formula
\begin{equation}
I_{nm}=e^{-\frac{1}{4} k^2} \, i^{n+m} \, (\frac{n!m!}{2^{n+m}})^{\frac{1}{2}}\, h_{nm}
\label{formula}
\end{equation}
where $h_{nm}$ satisfies the recursion equation
\begin{equation}
h_{nm}= (kh_{n-1,m}-2h_{n-1,m-1})/n 
\label{recursion}
\end{equation}
with the initial conditions $h_{00}=1$ and $h_{nm}=0$ if either of the
indices is negative. We proved formulas~(\ref{formula})~and~(\ref{recursion})
using generating function methods. The recursion formula is efficient
but numerically unstable. Therefore high precision arithmetics had to be used.

Since Fourier transform $\mathcal F$ is a unitary transform on
the space of square integrable functions and since Hermite functions
are eigenfunctions of $\mathcal F$, $\mathcal F \psi_n = i^n \psi_n $, 
we can also use the above formulas to calculate the convolutions
\begin{equation*}
\int_{-\infty}^{\infty} \psi_n(x)\psi_m(x-d) \, dx = i^{n+m}I_{nm}(d).
\end{equation*}
Apart from being interesting in their own right, these convolutions
give the matrix elements of the change of basis from $(\psi_n(x))$
to the translated basis $(\psi_n(x-d))$, which is exactly what 
was needed to transform the state matrix from the eigenstate basis 
of $H$ to that of $H^d$. Hence formulas~(\ref{formula})~and~(\ref{recursion})
were in double use in the implementation. 

For some values of the parameters it was beneficial to use the scaled
Hermite functions $(\sqrt{\lambda} \psi_n(\lambda x))$ with $\lambda > 1$
in the diagonalization to improve resolution. Even in this case the matrix elements needed
are easily derived from the above formulas. In the simulations we used less than 3000 basis
vectors and calculated typically a few hundred eigenstates. This was probably not enough for all
the cases but at this point our computer resources became a limiting factor. 

\section{Results} \label{simulation}

\begin{figure}
   \centering
   \includegraphics[totalheight=5truecm]{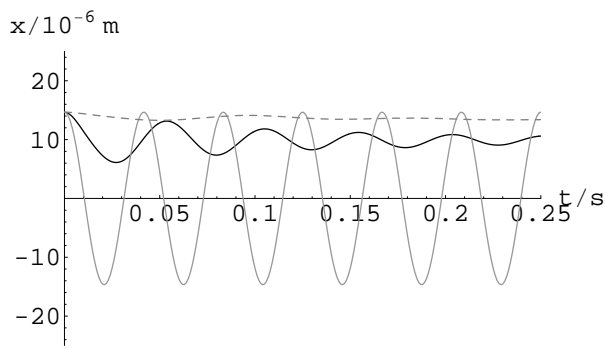}
   \caption{Time evolution of the center of mass of the fermion cloud at
            the lattice height $s=0$ (grey line), $s=3$ (black line), and
            $s=8$ (dashed line).} 
  \label{damped}
\end{figure}

\begin{figure}
   \centering
   \includegraphics[totalheight=5truecm]{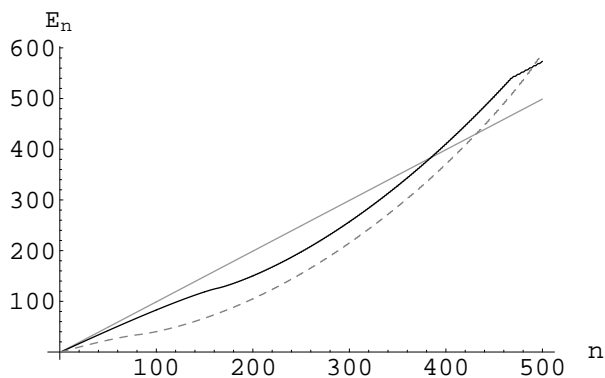}
   \caption{Energy spectrum of a single atom in the combined one-dimensional
    potential at the lattice height $s=0$ (grey line), $s=3$ (black line), and
    $s=8$ (dashed line). The energy unit is the energy quantum of the axial
     harmonic oscillator.  The ground state energies are set to zero. A transition 
   from a linear spectrum (harmonic oscillator) to a quadratic one occurs when the
   lattice is imposed. The quadratic part of the spectrum corresponds to localized
 states. Non-localized states appear again for higher energies, see the black line which becomes linear for $n>450$.}
    \label{spectra}
\end{figure}

We used the above methods to simulate the motion of the free
Fermi gas of potassium atoms in an optical lattice confined in a magnetic
potential and compared them to the experimental results described in~\cite{Pezze,modugno1,modugno2}. The average temperature of the gas in the experiment described 
in~\cite{Pezze}
was $T = 90$ nK, the lattice wavelength $\lambda = 863$ nm, the axial and radial 
frequencies $\omega_a=2\pi\times 24$ s$^{-1}$ and
$\omega_r = 2\pi \times 275$ s$^{-1}$,
respectively, and the average atom number $25000$.
Recall that $k=4\pi/\lambda$.
The depth $U=2A=sE_R$ of the lattice given in the units of recoil
energy $E_R=h²/2m\lambda^2$ was varied. Initially the trap was replaced by
$15$ $\mu$m to excite oscillations. The results of the simulation are shown in Figure~\ref{damped}.
The energies of the lowest lying single-particle eigenstates are also plotted
against the ordinal number in Figure~\ref{spectra} to give some qualitative insight
to the system.

In the absence of the lattice potential ($s=0$) 
the system is reduced to a harmonic oscillator and the average position
oscillates freely showing the characteristic dipole oscillation in
a harmonic potential. None of the eigenstates is localized and the corresponding
energy spectrum is linear. The energy quantum or the difference between the 
consecutive eigenvalues gives the frequency of the oscillations. 

When the height of the lattice is increased to $s=3$, both localized and
non-localized eigenstates play a role. The oscillations of the system are
damped, which can be understood as follows.
According to the formula~(\ref{timeevolution}), the center of mass of
the cloud is a sum periodic terms whose frequencies are given by the 
differences $(\varepsilon_n-\varepsilon_m)/\hbar$ between 
the eigenvalues $\varepsilon_n$ of the dynamical Hamiltonian. When the eigenvalues
are not evenly spaced, destructive interference causes dephasing. 

\begin{figure}
   \centering
   \includegraphics[width=12cm]{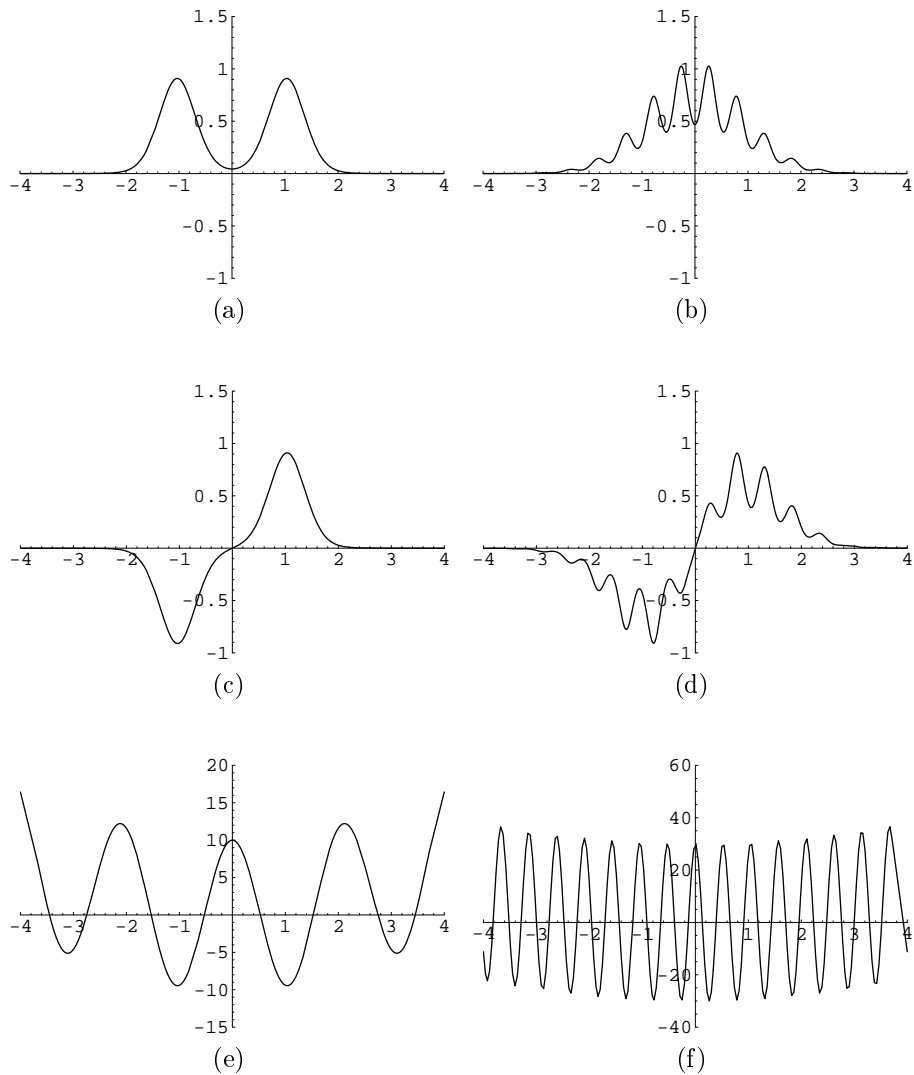}
   \caption{The two eigenstates (a) and (c) of the lowest energy associated to the potential (e) 
with parameters $k=3$ and $A=10$ form a quasidegenerate pair. In a superposition of such states, the particle is localized in one of the wells of the potential (e). Relating to Figure 2, such states correspond to a doubly degenerate pair of eigenvalues in a quadratic spectrum. 
The eigenstates (b) and (d) associated to the
potential (f) with parameters $k=12$ and $A=30$ are not localized. They resemble
the eigenstates of the harmonic oscillator, modulated by the lattice potential. They correspond to a linear spectrum and their energies are not degenerate.}
   \label{eigenstates}
\end{figure}

Increasing the lattice height also increases 
the offset of the dipolar oscillations since more and more particles become
localized. 
When the height of the lattice is increased  to $s = 8$, localized
eigenstates dominate. The system is almost frozen to its initial position.
The energy spectrum becomes almost quadratic. The spectrum corresponds to states that 
are localized 
in individual lattice potential sites. Neighbouring sites are shifted in energy 
according to the 
superimposed harmonic potential, therefore the quadratic form of the spectrum 
(difference in energy of two
localized particles depends essentially only on their locations in the lattice).
Actually each energy is doubly
degenerate (the symmetric and the antisymmetric states formed of lattice sites at 
opposite sides of the center) 
which can be resolved by a closer look at the spectrum. Therefore, to be accurate,
localized particles correspond
to a superpositions of these doubly degenerate states. Representative examples of
localized and non-localized eigenstates are shown in Figure~\ref{eigenstates}.

It is noteworthy that in the intermediary region lowest-lying states are
not localized. This is indicated by the linear parts in the spectra near
origin. Hence the atoms do not tend to localize when the system is cooled down
to ultra-low temperatures, on the contrary. Note that also higher energy states may be unlocalized, i.e. have a linear spectrum, as is shown for one set of parameter values in Figure~\ref{spectra}. This is associated with the existence of higher bands of the optical lattice potential. Such states may affect the localization and oscillation behaviour of the gas significantly in higher temperatures.       

The results are in very good qualitative and quantitative agreement both with 
the experiment and the semiclassical approximation presented in~\cite{Pezze}.
We compared our simulations also to two earlier experiments~\cite{modugno1,modugno2} 
by Modugno et al. In these cases the experimentalists observed oscillations with considerably 
larger amplitude and higher frequency than what was predicted by the simulations.
We believe that this was simply because the physical parameters used in these experiments were
much more demanding for our numerical method and would have required more
eigenstates to be calculated than was currently possible. The larger the displacement and the 
particle number, and
the higher the temperature, the more eigenstates are needed for reliable results.
The computer resources at our disposal were rather modest so that we have hardly pushed the
method to its limits. 

\section{Conclusions}

We have analyzed the energy spectrum and motion of a Fermi gas in a combined potential.
The harmonic potential alone has a linear spectrum and shows dipole oscillations of the gas 
when the trap center is 
displaced. Introducing the periodic potential transforms the spectrum gradually into
 quadratic one and the 
gas becomes localized. Intermediate cases show damped oscillations around the displaced
center position. Our approach is complementary to the semiclassical analysis presented in \cite{Pezze,yksi}, and
is in good qualitative agreement with the experiments. For the parameter values which are not too challenging from
the numerical point of view the results agree also quantitatively with both the experiments and the semiclassical
analysis.

It is important to note that the quantum
transport in the combined harmonic and periodic potential is qualitatively rather different from the Bloch 
oscillation or Wannier-Stark ladder behaviour in tilted periodic potentials.
The two cases approach each other only when the gas moves in a region where the harmonic trap can be 
approximated by a linear potential.
This requires a much shallower trap than that used in the experiments \cite{modugno1,modugno2,Pezze}. 
Understanding the behaviour of the non-interacting Fermi gas sets the basis for investigating effects of 
interactions \cite{modugno2,Stringari} and eventually superfluidity \cite{Mirta,muu}. 

\ack{We thank G.~Modugno, H.~Ott and M.~Inguscio for useful discussions. We acknowledge the 
financial support from the Academy of Finland (Project Nos.\ 53903, 205470), Emil Aaltonen foundation 
and European Commission IST-2001-38877 (QUPRODIS).}

\appendix
\section*{Appendix}
\setcounter{section}{1}

In this appendix we formulate the reduction from many-particle to single-particle 
state in the language of symplectic geometry. This formulation
may be helpful in using the reduction in a more systematic way and perhaps 
for more general purposes. 

Let us first recall some concepts of classical mechanics using a point-like particle
moving in three-dimensional space as an example. The configuration space of the
particle is just $C=\mathbb{R}^3$ whereas the coordinates of phase space
 $M\simeq \mathbb{R}^3\times \mathbb{R}^3$
include the momentum $\mathbf{p}$ as well as the position $\mathbf{x}$. In
mathematical terms, the phase space is the cotangent space $T^*C$ of the configuration
space and it comes equipped with a canonical non-degenerate two-form
\begin{equation*}
\omega = \sum_{k=1}^3 dp^k \wedge dx^k,
\end{equation*}   
which is known as the symplectic form while the pair $(M,\omega)$ is
referred to as a symplectic manifold.  The symmetries of the phase space
are the diffeomorphisms of $M$ preserving $\omega$. Accordingly, infinitesimal
symmetries are the vector fields $A$ on $M$ whose Lie derivative satisfy
$L_A\omega = 0$. Then each infinitesimal symmetry generates a one-parameter
group of symmetries at least locally, where the correspondence is given by the flow
of the vector field.  

Under some mild conditions we can associate a classical observable or   
a function $H_A$ on the phase space $M$ to an infinitesimal symmetry $A$.
The observable is defined by the equation 
\begin{equation*}
\omega(A,B) = dH_A(B)
\end{equation*}
which is required to be valid for all vector fields $B$ on $M$. In particular,
since the time-evolution of a classical mechanical system preserves the symplectic
form, it is generated by an observable, which is know as the Hamiltonian. 
In the example above it is clear that both translations and rotations are symmetries
of the phase space. An infinitesimal translation is just a three-vector
$\mathbf{t} \in \mathbb{R}^3$, and the corresponding observable is 
\begin{equation*}
H_{\mathbf{t}}(\mathbf{x},\mathbf{p})=\mathbf{p} \cdot \mathbf{t}. 
\end{equation*}
Similarly, an infinitesimal rotation can be given by a three-vector $\mathbf{r}$ 
as $\mathbf{r} \cdot \mathbf{R}$
where the components of $\mathbf{R}$ are the infinitesimal
rotations about the coordinate axes. The associated classical observable is then
\begin{equation*}
H_{\mathbf{r}}(\mathbf{x},\mathbf{p})=(\mathbf{x} \times \mathbf{p}) \cdot \mathbf{r}. 
\end{equation*}
The general pattern emerging is that given a Lie algebra $\mathfrak {g}$ of
infinitesimal symmetries, the collection of associated observables is most
conveniently expressed as a map $\rho$ from the phase space $M$ to the dual $\mathfrak{g}^*$,
that is, the space of linear maps $\mathfrak{g} \to \mathbb{R}$. In this way
we associated the momentum $\mathbf{p}$ to the translations and the impulse moment
$\mathbf{x} \times \mathbf{p}$ to the rotations. Therefore the map  
$\rho: M \to \mathfrak{g}^* $ defined by the equation
\begin{equation*}
(\rho(x),A)=H_A(x)
\end{equation*} 
for all $A \in \mathfrak{g}$ and $x\in M$ is known as the moment map.
The Lie group $G$ corresponding to the Lie algebra $\mathfrak{g}$ acts naturally
on both $\mathfrak{g}$ and $\mathfrak{g}^*$, and the moment map is equivariant in
the sense that 
$\rho(gx) = g\rho(x)
$for all $g$ in $G$.

Returning to quantum mechanics,
let $\mathcal{H}$ be a Hilbert space, on which a compact group $G$ acts unitarily. 
This is the general framework for a quantum mechanical system with 
the symmetry group $G$. Now the projective space
\begin{equation*}
\mathbb{P}(\mathcal{H})=\lbrace v \in \mathcal{H} \: \vert \: \Vert v \Vert = 1 
\rbrace \: / \: U(1)
\end{equation*}
of pure states is canonically a symplectic manifold, on which $G$ acts preserving the 
symplectic form. Hence from the point of view of symplectic geometry, all that we
have said previously applies.

Let $\mathfrak{g}$ be the Lie algebra of $G$.
The observable associated to an infinitesimal symmetry 
$A \in \mathfrak{g}$ is calculated to be the expectation value
$H_A(v) = \langle v \vert \hat{A}  \vert v \rangle$ where
$\hat {A}$ stands for the representation of $A$ in $\mathcal {H}$.  
There is also the moment map 
\begin{equation*}
\rho: \mathbb{P}(\mathcal{H}) \to \mathfrak{g}^* 
\end{equation*}
from the projective space to the dual of $\mathfrak{g}$
defined by the equation 
\begin{equation*}
H_A(v)=( \rho(v), A ).
\end{equation*}
Hence the expectation value of $\hat {A}$ in the vector
state $v$ is given by $\langle \hat{A} \rangle = ( \rho(v), A )$.
We can extend the moment map to any state $\Phi$ by linearity.
Then the moment map image $\rho(\Phi)$ is a kind of reduced state 
which can be used to calculate expectation values 
for all observables in $\mathfrak{g}$. The point is that $\rho(\Phi)$ may
be a much simpler object than the state $\Phi$ we started with.
Furthermore, the equivariance of the moment map means that
\begin{equation*}
\rho(\hat{U}\Phi \hat{U}^{-1})=U\rho(\Phi)U^{-1}
\end{equation*}
for every element
$U$ of $G$. In particular,
if the Hamiltonian $H$ itself belongs to $\mathfrak{g}$, then the time evolution of
the reduced state $\rho(\Phi)$ is given by the usual Schrödinger equation. 

In the application we have in mind
$G$ is the group of unitary transformations of the single-particle
Hilbert space $\mathcal V$ and $\mathfrak{g}$ may be identified with the
single-particle observables or the Hermitian operators on $\mathcal V$.
Using the pairing $( B, A ) = \Tr (BA)$ we may 
identify $\mathfrak{g}$ with its dual. Let $\mathcal{H}$ be the
fermionic Fock space associated to $\mathcal V$. Then $G$ acts naturally on
$\mathcal{H}$ and by the general theory
\begin{equation}
\langle \hat{A} \rangle = \Tr (\Phi \hat{A}) =  \Tr (\rho(\Phi)A)
\label{expectation}
\end{equation} 
for any many-particle state $\Phi$ and single-particle observable $A$.
It is easy to check that $\rho(\Phi)$ is a positive operator 
with trace
\begin{equation*}
\Tr{\rho(\Phi)} = \Tr (\rho(\Phi)I) = \Tr (\Phi \hat {I})=
\Tr (\Phi N)= \langle N \rangle
\end{equation*}                 
since the number operator $N$ is indeed the representation of $I \in \mathfrak{g}$
in the fermionic Fock space.

In conclusion, we can represent any many-particle state $\Phi$
by a single-particle state $\rho(\Phi)$ when calculating expectation
values of single-particle observables. It is easy to calculate
$\rho(\Phi)$ explicitly in a given basis $(v_n)$
of $\mathcal{V}$ using equation~\ref{expectation}. In terms of the 
annihilation and creation operators associated to this basis, 
the matrix elements of $\rho(\Phi)$  are 
\begin{equation*}
\rho(\Phi)_{nm}
=\Tr (\Phi  a_m^\dagger a_n ) = \langle a_m^\dagger a_n \rangle. 
\end{equation*} 
Applying the above procedure to the grand canonical ensemble~(\ref{gce})
yields 
\begin{equation*}
\rho(\Phi(0))_{nm} = 
\frac{e^{\beta(\mu-\epsilon_n)} \prod_{n\neq k=1}^{\infty} 
        e^{\beta(\mu-\epsilon_k)}}
     { \prod_{k=1}^{\infty} e^{\beta(\mu-\epsilon_k)}} \delta_{nm}
=f(\epsilon_n-\mu) \delta_{nm}  ,
\end{equation*} 
where 
\begin{equation*}
f(x) = \frac {1}{1+e^{\beta x}}
\end{equation*}
is the Fermi-Dirac distribution which is usually derived using the partition
function approach. The preceeding discussion was not just for the
sake of an appealing generalization but also to show that the reduction
to the single-particle state formalism gives not only the energy distribution
but a state matrix obeying the Schr\"odinger equation.

\end{document}